# The Power-Weakness Ratios (PWR) as a Journal Indicator: Testing the "Tournaments" Metaphor in Citation Impact Studies

Loet Leydesdorff,[a]* Wouter de Nooy,[b] and Lutz Bornmann[c]

* corresponding author
[a] University of Amsterdam, Amsterdam School of Communication Research, P.O. Box 15793 1001 NG Amsterdam, The Netherlands; email: loet@leydesdorff.net

[b] University of Amsterdam, Amsterdam School of Communication Research, P.O. Box 15793 1001 NG Amsterdam, The Netherlands; email: w.denooy@uva.nl

[c] Division for Science and Innovation Studies, Administrative Headquarters of the Max Planck Society, Hofgartenstr. 8, 80539 Munich, Germany; email: bornmann@gv.mpg.de

**Abstract**

Ramanujacharyulu's (1964) Power-Weakness Ratio (PWR) measures impact by recursively multiplying the citation matrix by itself until convergence is reached in both the cited and citing dimensions; the quotient of these values is defined as PWR, whereby "cited" is considered as power and "citing" as weakness. Analytically, PWR is an attractive candidate for measuring journal impact because of its symmetrical handling of the rows and columns in the asymmetrical citation matrix, its recursive algorithm, and its mathematical elegance. In this study, PWR is discussed and critically assessed in relation to other size-independent recursive metrics. A test using the set of 83 journals in "information and library science" (according to the Web-of-Science categorization) converged, but did not provide interpretable results. Further decomposition of this set into homogeneous sub-graphs shows that—like most other journal indicators—PWR can perhaps be used within homogeneous sets, but not across citation communities.

**Introduction**

Ramanujacharyulu (1964) provided a graph-theoretical algorithm to select the winner of a tournament on the basis of the total scores of all the matches, whereby both gains and losses are taken into consideration. Prathap & Nishy (in preparation) proposed to use this power-weakness ratio (PWR) for citation analysis and journal ranking. Analytically, PWR is an attractive candidate for measuring journal impact because of its symmetrical handling of the rows and columns in the asymmetrical citation matrix, its recursive algorithm (which it shares with other journal indicators), and its mathematical elegance. However, Ramanujacharyulu (1964) developed the algorithm for scoring tournaments (Prathap, 2014). Can journal competitions be compared to tournaments? Note that journals compete in incomplete tournaments; in a round-robin tournament, all the teams are completely connected. If one team wins, the other loses. This constraint is not valid for journals.

Prathap & Nishy (in preparation) proposed to explore the power-weakness ratio (PWR) for citation analysis and journal ranking, since, in their opinion, this measure can be expected to improve on the influence weights proposed by Pinski and Narin (1976), the Eigenfactor and Article Influence Scores (Bergstrom, 2007; West *et al*., 2010), the PageRank (Brin & Page, 2001), and the Hubs-and-Authorities thesis (Kleinberg, 1999) on the Web Hypertext Induced Topic Search (HITS). PWR shares with these algorithms the ambition to develop a size-independent metric based on recursion in the evaluation of the accumulated advantages (Price, 1976). Unlike these other measures, in PWR the disadvantages are appreciated equally with the advantages; the "power" (gains) is divided by the "weakness" (losses). In studies of sporting tournament (e.g., crickets), the ranking using PWR was found to outperform other rankings (Prathap, 2014).

In this study, we respond to this proposal in considerable detail by testing PWR empirically in the citation matrix of 83 journals assigned to the Web-of-Science (WoS) category "information and library science" (LIS) in the Journal Citation Reports 2013 of Thomson Reuters. This set is known to be heterogeneous (Leydesdorff & Bornmann, 2016; Waltman *et al*. 2011a): in addition to a major divide between a set of library and information science journals (e.g., *JASIST*) and a somewhat smaller group of journals focusing on management information systems (e.g., *MIS Quarterly*), a number of journals are not firmly related to the set, and one can further distinguish a relatively small group of bibliometrics journals within this representation of the library and information sciences (Milojević & Leydesdorff, 2013).

We focus the discussion first on the entire set and then decompose into two sub-graphs of journals: (1) seven journals which cited *JASIST* at least one hundred times during 2013, and (2) nine journals that cited *MIS Quarterly* a hundred or more times. Furthermore, we study the effect of combining these two subsets into an obviously heterogeneous set of (7 + 9 =) 16 journals. The conclusion will be that the relatively homogeneous subsets converge quickly, but in the case of the heterogeneous set, PWR convergence is more slowly. At the level of the total set of 83 journals, convergence was reached, but the results were not interpretable.

The results indicate that one is not allowed to compare impact across borders between homogenous sets because citation impacts can be expected to mean something different in other systems of reference. More recently, Todeschini *et al.* (2015) proposed a weighted variant of PWR ("wPWR") for situations where the criteria can have different meanings and relevance. However, we have no instruments for weighting citations across disciplines and the borders of specialties in terms of journal sets are fuzzy and not given (Leydesdorff, 2006).

In other words, scholarly publishing can perhaps be considered in terms of tournaments, but only within specific domains. Journals do not necessarily compete in terms of citations across domains. Citation can be considered as a non-zero game: if one player wins, the other does not necessarily lose, and thus the problem is not constrained, as it is in tournaments. Since there are no precise definitions of homogeneous sets, interdisciplinary research can be at risk, while the competition is intellectually organized mainly within core set(s) (Rafols *et al.*, 2012).

**Recursive and size-independent algorithms for impact measurement**

Among the journal indicators, the first distinction is between size-dependent and size-independent ones (De Visscher, 2010 and 2011; Leydesdorff, 2009; Pinski & Narin, 1976). The numbers of publications and citations, for example, are size-dependent indicators: large journals (e.g., *PNAS*, *PLoS ONE*) contain more publications and therefore, *ceteris paribus,* can be expected to contain more references and be more frequently cited.

Garfield & Sher (1963) first introduced the journal impact factor (JIF) as a size-independent measure of journal influence. In the case of JIF, the number of citations (e.g., in year $t$) is divided by the number of publications (e.g., in the years $t$-1 and $t$-2). More generally, the ratio of citations over publications (*C/P*) is a size-independent indicator. Pinksy & Narin (1976; cf. Narin, 1976) proposed to improve on Garfield's (1972) JIF by normalizing citations not by the number of publications, but by the aggregated number of ("citing") references in the articles during the publication window of the citation analysis. Yanovski (1981, at p. 229) called this quotient between citations and references the "citation factor." The citation factor was further elaborated into the "Reference Return Ratio" by Nicolaisen and Frandsen (2008). In the numerator, however, Pinski & Narin (1976) used a recursive algorithm similar to the one used below for the numerator and denominator of PWR. This example of an indicator based on a recursively converging algorithm was later followed with modifications by the above-mentioned authors of PageRank, HITS, Eigenfactor, and the Scimago Journal Ranking (SJR; Guerrero-Bote *et al*., 2012).

"Eigenfactor", for example, can as a numerator be divided by the number of articles in the set in order to generate the so-called "article influence score" (West *et al*., 2010; cf. Yan & Ding, 2010). Using Ramanujacharyulu's (1964) PWR algorithm, however, the same recursive algorithm is applied in the cited-direction to the numerator and in the citing-direction to the denominator. "Being cited" is thus considered as contributing to "power" whereas citing is considered as "weakness" in the sense of being influenced. Different from tournaments, however, the "winning" of one player does not imply "losing' by another in the case of citations. Let us assume that these are cultural metaphors—we return to this in the discussion—and continue first to investigate the properties of the indicator empirically. For a mathematical elaboration, the reader is referred to Todeschini *et al*. (2015).[1]

---

[1] In another context, Opthof & Leydesdorff (2010) noted that indicators based on the ratio between two numbers (such as "rates of averages") are no longer amenable to statistical analysis such as significance testing of differences

**The Power-Weakness Ratio (PWR)**

Let $\mathbf{Z}$ be the cited-citing journal matrix. If the entries are read row-wise, then for a journal in row $i$, an entry such as $\mathbf{Z_{ij}}$ denotes the citations from journal $j$ in the citation window (say 2013) to articles published in journal $i$ during the publications window (say 2011–2012); in social-network analysis these are considered the in-coming links. When the matrix is read column-wise; now for the journal in column $j$, the entry $\mathbf{Z_{ij}}$ signifies the references from journal $j$ in the citation window (2013) to articles published in journal $i$ during the publications window (2011–2012). In social-network analysis these are considered the out-going links.[2]

Using graph theory, $\mathbf{Z} = [\mathbf{Z_{ij}}]$ is the notation of the matrix associated with the graph. Many properties of such matrices are known, and it can be raised indefinitely to the $k^{th}$ power, i.e., $\mathbf{Z}^k$. The Eigenfactor, for example, is a recursive iteration that raises $\mathbf{Z}$ to an order where convergence is obtained for what is effectively the weighted value of the total citations (Yan & Ding, 2010). One can find a value $p_i(k)$ for each journal; this can be called the iterated power of order $k$ of the journal $i$ "to be cited". The recursive procedure for formalizing the computation of $p_i(k)$ is given in graph theoretic terms in Ramanujacharyulu (1964). An algorithmic implementation using the Stodola method of iteration is provided by Dong (1977).

---

among the resulting values (Gingras & Larivière, 2011). More recently, other indicators based on comparing observed with expected values have also been introduced (e.g., *MNCS* by Waltman *et al.*, 2011b; *I3* by Leydesdorff *et al.*, 2012; cf. Leydesdorff *et al.*, 2011).

[2] In social network analysis, the matrix is usually transposed so that action ("citing") is considered as the row vector.

One can carry out the same operations column-wise by using the transposed matrix $\mathbf{Z}^T$ and then proceeding row-wise among these transposed elements in the same recursive and iterative manner as above. Again, for each journal one can find a value $w_i(k)$, which can be considered the iterated weakness of the order $k$ of the journal $i$ "to be influenced by." The empirical question remains of whether both $p_i(k)$ and $w_i(k)$ converge for $k \to \infty$. If $k \to \infty$ converges, one obtains the converged power-weakness ratio $r_i(k) = p_i(k)/w_i(k)$.

From this perspective, a journal is considered powerful when it is cited by other powerful journals and is weakened when it cites other weaker journals. This dual logic of PWR is similar to the Hubs and Authorities thesis of the Web Hypertext Induced Topic Search (HITS), a ranking method of webpages proposed by Kleinberg (1999); but with one major difference. In the HITS paradigm as applied to a bibliometric context, good authorities would be those journals that are cited by good hubs, and good hubs the journals that cite good authorities. Using PWR, however, good authorities are journals that are cited by good authorities and weak hubs are journals that cite weak hubs. Using CheiRank (e.g., Zhirov *et al*., 2010), the two dimensions of power and weakness can also be considered as *x*- and *y*-axes in the construction of two-dimensional rankings. A review of ranking techniques using PageRank-type recursive procedures is provided by Franceschet (2011).

We study the effectiveness of the proposed indicator using journal ecosystems drawn from the Library and Information Science set of the Web of Science (83 journals) as an example. Two local ecosystems (sub-graphs) are isolated from this larger scientific network and the cross-citation behaviour within each sub-graph is analyzed. Can the indicator be a measure of the standing of each journal in the cross-citation activity within a sub-graph that is more finely-grained than, for example, the journal impact factor or other indicators defined at the level of the total set? We will also compare with the Scimago Journal Ranking (SJR) because this indicator uses a recursive algorithm similar to PageRank.

**Methods**

One can perform the recursive matrix multiplication to the power of a matrix in a spreadsheet program such as Excel. Excel 2010 additionally provides the function MMult() for matrix multiplications, but this function operates with a maximum of 5,460 cells (or $n \leq 73$). Matrix multiplications are computationally intensive. However, the network analysis and visualization program Pajek (de Nooy *et al*., 2011) can also be used for matrix multiplication in the case of large sets. We used Pajek to compute PWRs for the full set of 83 journals with the LIS category, and Excel for the computation in the case of the two smaller sub-graphs: (1) *JASIST+* contains the seven journals that cite *JASIST* more than 100 times in 2012; and (2) *MIS Quart+* the nine journals citing this journal to the same extent.

A macro (PWR.MCR) for Pajek is provided at http://www.leydesdorff.net/pwr/pwr.mcr which generates PWR values for $k = 1$ to $k = 20$ as vectors from a one-mode (asymmetrical) citation matrix with an equal number of rows and columns. Similarly, the Excel file for the *JASIST+* set can be retrieved from http://www.leydesdorff.net/pwr/jasist.xlsx. Using the function MMult() in Excel, one can replace cell J4 with "=MMULT($B4:$H4,I$4:I$10)", etc., *mutatis mutandis* (available at http://www.leydesdorff.net/pwr/mmult.xlsx).[3] The results of the various methods are similar except for rounding errors caused by how one deals with the main diagonal.

The values on the main diagonal represent within-journal self-citations. One can argue that self-citations should not be included in subsets since the number of self-citations is global: it remains the same in the total set and in subsets, and therefore may distort subsets (Narin & Pinsky, 1976, p. 302; cf. Price, 1981, p. 62). In a second sheet "without self-citations", we show that in this case the effects are marginally different. In Appendices 1 and 2, the procedures for using Pajek or Excel, respectively, are specified in more detail.

**Results**

*a. The LIS set (83 journals)*

Among the 83 journals assigned to the journal category "information and library science" by Thomson Reuters, one is not cited within this set and four journals do not cite any of these journals. Seventy-five of the 83 journals are part of a single strong component, so they are mutually reachable directly or indirectly; the remaining eight journals include journals that are only cited by other journals, only cite other journals, or are neither cited nor citing. Note that journals that are cited but not citing obtain (very) high PWR scores because their weakness score in the denominator is minimal;[4] however, these journals do not affect the PWR scores of the other journals. Probably, one is well advised to limit the applications of PWR to strong components.

---
[3] In Excel, we use the so-called Stodola method, which simplifies the computation (e.g., Dong, 1977). However, upon extension to the full set and $k = 20$, the results are similar to those obtained using Pajek except for rounding errors.

**Table 1**: Fifteen journals ranked highest on PWR among 83 LIS journals.

| Abbreviation of journal name | PWR |
|---|---:|
| Int J Comp-Supp Coll | 59.52 |
| MIS Q Exec | 15.62 |
| Inform Syst Res | 11.31 |
| Libr Quart | 8.84 |
| MIS Quart | 6.96 |
| J Manage Inform Syst | 6.15 |
| J Med Libr Assoc | 5.53 |
| Inform Manage-Amster | 5.01 |
| J Am Med Inform Assn | 4.40 |
| Inform Organ-UK | 4.20 |
| J Acad Libr | 3.57 |
| J Inf Technol | 3.38 |
| J Health Commun | 3.15 |
| Inform Soc | 3.09 |
| Aust Acad Res Libr | 2.90 |

Table 1 lists ranked PWR values for 15 of the 75 journals in the central component after 20 iterations (after removing the four non-citing journals). *JASIST*, for example, follows with a much lower PWR value of 1.45.

All PWR values were stable at $k = 20$. However, it is difficult at this stage to say whether this ranking provides a meaningful measure of journal impact, which is one measure of the bibliometric performance indicators of a journal. PWR is a candidate size-independent measure that can play this role. Our results can be considered as a test of this hypothesis. In our opinion,

---
[4] The weakness score in this case is determined by the number of self-citations on the main diagonal and otherwise zero.

PWR failed as an indicator of overall journal standing since we were not able to provide the results in Table 1 with an interpretation. Note that the Pajek macro can handle large network data (e.g., the complete JCR).

**Decomposition of the LIS set**

As noted above, some journals never cited another journal in this set and one journal never received any citations from the other journals in the set. For analytical reasons, PWR would be zero in the latter case and may go to infinity in the former. However, a structural analysis shows that there are two main sub-graphs in this set. These can, for example, be visualized by using the cosine values between the citing patterns of 78 (of the 83) journals (Figure 1).

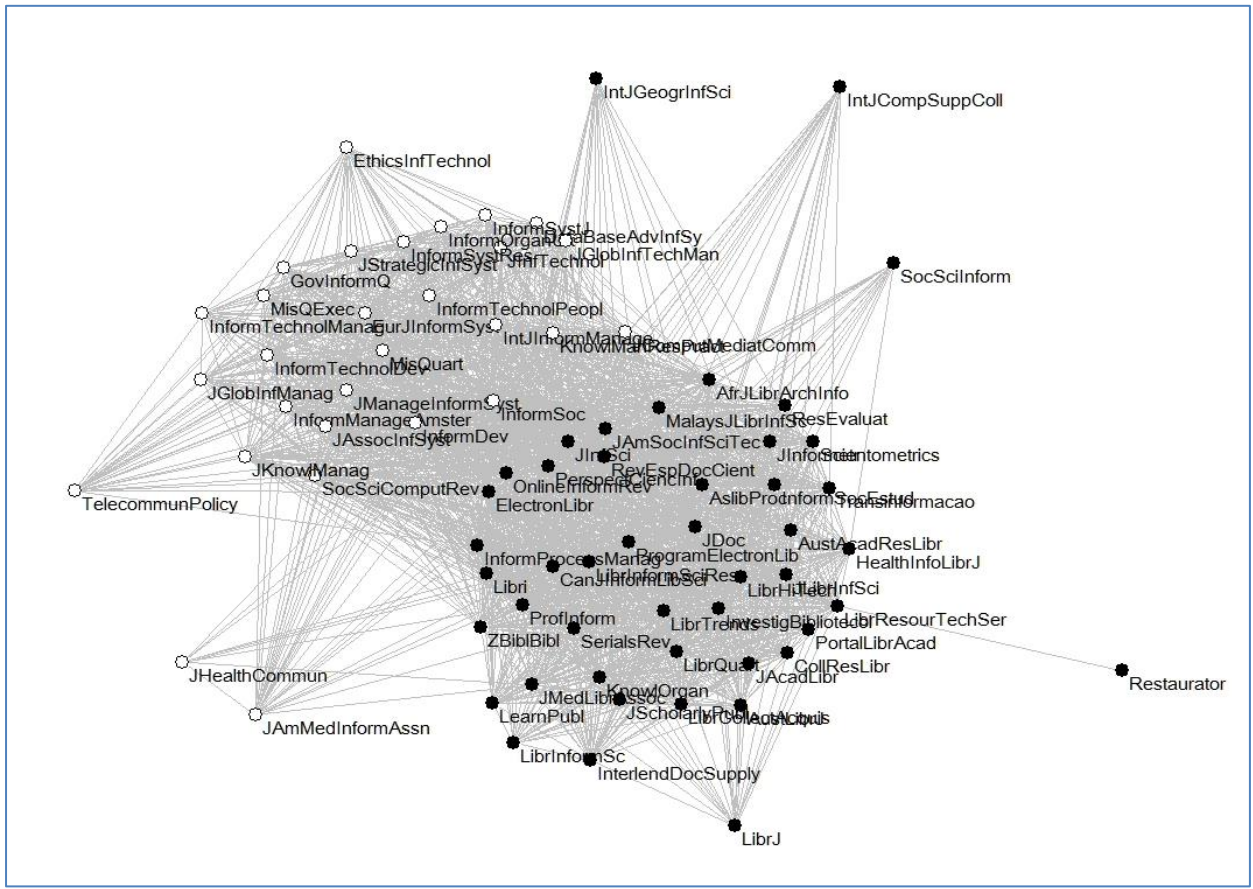

**Figure 1**: Two groups of journals within the WoS category "information and library science"; cosine > 0.01; $Q$ = 0.359; Blondel *et al.* (2008); Kamada & Kawai (1989) used for the visualization.

Using the Louvain algorithm for the decomposition of this cosine-normalized matrix, 40 of these journals are assigned to partition 1 (LIS - library & information science) and 38 to partition 2 (MIS - management information systems; cf. Leydesdorff & Bornmann, 2016). From these two subsets, we further analysed two ecosystems which were selected because they are well-connected homogeneous sets.

**Table 2**. The two homogeneous journal sub-graphs chosen for further analysis, and their abbreviated journal names.

| Sub-graph | Abbreviated Name |
|---|---|
| **JASIST+** | INFORM PROCESS MANAG |
| | J DOC |
| | J AM SOC INF SCI TEC |
| | J INF SCI |
| | SCIENTOMETRICS |
| | J INFORMETR |
| | INFORM RES |
| **MIS QUART+** | EUR J INFORM SYST |
| | INFORM MANAGE-AMSTER |
| | J ASSOC INF SYST |
| | J INF TECHNOL |
| | J MANAGE INFORM SYST |
| | J STRATEGIC INF SYST |
| | MIS QUART |
| | INFORM SYST RES |
| | INT J INFORM MANAGE |

Table 2 shows the two homogeneous journal ecosystems chosen for further study (using abbreviated journal names). The JASIST+ set comprises seven journals, all of which have cited

*JASIST* at least 100 times and come from the LIS partition. The MIS QUART+ set is similarly a set of nine journals strongly connected to one another within the MIS partition.[5] Finally, we shall combine the JASIST+ and MIS QUART+ sets into a set of 16 journals so that inhomogeneity is built into this arrangement.

For each ecosystem, we take the year 2013 as the citation window and the publication window as all years (total cites). Since all journals are well connected within the sub-graphs, there are no dangling nodes (where the journals are cited within the ecosystem but hardly cite any other journal in the same system). Using PWR, no damping or normalization (as is used in the PageRank approach) is required: one can use the cross-citation matrix without further tuning of parameters. In each case, when $k = 1,$ one obtains the raw or non-recursive value of impact, and when the iteration is continued to higher orders of $k$ as $k \to \infty$ convergence of the recursive power-weakness ratios was found in both sets.

**Table 3**. Citation matrix $\mathbf{Z}$ for the JASIST+ set of seven journals.

| *Citing* <br> *Cited* | INFORM PROCESS MANAG | JASIST | J INF SCI | SCIENTO METRICS | INFORM RES | J DOC | J INFORM ETR |
|---|---|---|---|---|---|---|---|
| INFORM PROCESS MANAG | 132 | 165 | 49 | 86 | 68 | 46 | 23 |
| JASIST | 120 | 756 | 107 | 495 | 189 | 139 | 319 |
| J INF SCI | 12 | 66 | 89 | 72 | 26 | 26 | 30 |
| SCIENTOMETRICS | 48 | 320 | 34 | 1542 | 13 | 25 | 552 |
| INFORM RES | 14 | 43 | 29 | 8 | 93 | 39 | 4 |
| J DOC | 26 | 96 | 44 | 69 | 128 | 108 | 29 |
| J INFORMETR | 29 | 91 | 2 | 269 | 4 | 3 | 302 |

---

[5] Unlike the JASIST+ set, the MIS Quart+ set is not a completely connected clique, since the *International Journal of Information Management* was not cited by articles in the *Journal of Information Technology* during 2013.

Table 3 shows the citation matrix $\mathbf{Z}$ for the JASIST+ set of 7 journals. The weakness matrix can be obtained as the transpose of this matrix, and the cases without self-citation are obtained by discarding the entries in the diagonal and replacing them with zeroes. In Table 4 we report the convergence of the size-independent power-weakness ratio $r$ with iteration number $k$ for the JASIST+ journals for the cases with and without self-citations. We see that this indicator can serve as a proxy for the relative qualities or specific impacts of the journals within this set.

**Table 4**. Convergence of PWR with iteration $k$ for the JASIST+ journals, with and without self-citations.

| With self-citations | | | | | | | |
|---|---|---|---|---|---|---|---|
| PWR r for k= | 1 | 2 | 3 | 4 | 5 | 6 | 7 |
| INFORM PROCESS MANAG | 1.49 | 1.72 | 1.76 | 1.76 | 1.76 | 1.76 | 1.77 |
| J DOC | 1.30 | 1.38 | 1.52 | 1.60 | 1.63 | 1.64 | 1.64 |
| JASIST | 1.38 | 1.48 | 1.51 | 1.53 | 1.53 | 1.53 | 1.54 |
| J INF SCI | 0.91 | 1.19 | 1.36 | 1.43 | 1.45 | 1.46 | 1.46 |
| SCIENTOMETRICS | 1.00 | 0.98 | 0.98 | 0.98 | 0.98 | 0.98 | 0.97 |
| J INFORMETR | 0.56 | 0.48 | 0.47 | 0.47 | 0.47 | 0.47 | 0.47 |
| INFORM RES | 0.44 | 0.37 | 0.39 | 0.40 | 0.41 | 0.41 | 0.41 |
| Without self-citations | | | | | | | |
| PWR r for k= | 1 | 2 | 3 | 4 | 5 | 6 | 7 |
| INFORM PROCESS MANAG | 1.76 | 1.93 | 1.75 | 1.80 | 1.78 | 1.79 | 1.79 |
| JASIST | 1.75 | 1.52 | 1.61 | 1.57 | 1.59 | 1.58 | 1.58 |
| J DOC | 1.41 | 1.46 | 1.46 | 1.48 | 1.47 | 1.48 | 1.48 |
| J INF SCI | 0.88 | 1.23 | 1.23 | 1.25 | 1.25 | 1.25 | 1.25 |
| SCIENTOMETRICS | 0.99 | 0.99 | 0.98 | 0.99 | 0.98 | 0.99 | 0.98 |
| J INFORMETR | 0.42 | 0.49 | 0.48 | 0.48 | 0.48 | 0.48 | 0.48 |
| INFORM RES | 0.32 | 0.43 | 0.41 | 0.42 | 0.42 | 0.42 | 0.42 |

Table 4 shows, among other things, that the inclusion of self-citations affects PWR values in this case only in the second decimal.

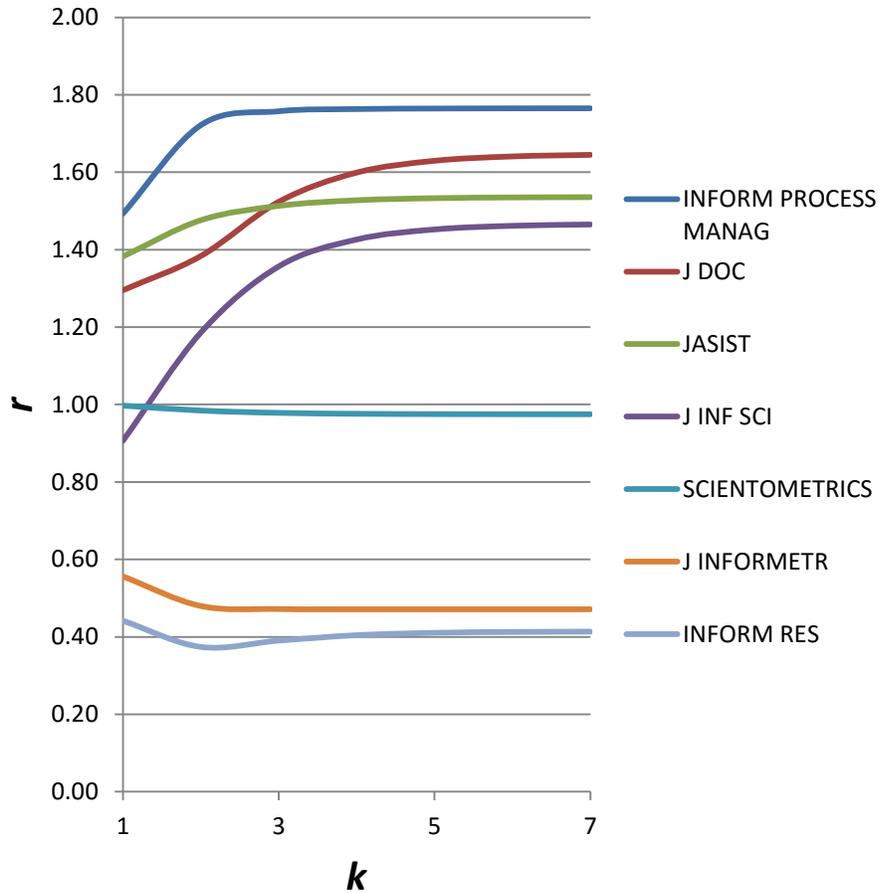

**Figure 2:** Convergence of PWR with iteration number $k$ for the seven JASIST+ journals for the case without self-citations.

Figure 2 graphically displays the convergence of PWR with iteration number $k$ for the JASIST+ set without self-citations. As noted, it may be meaningful to proceed with the case where self-citations are ignored. Analogously, Figure 3 shows the convergence of PWR for the MIS QUART+ set without self-citations. Again, within this homogeneous set rapid and stable convergence of the PWR values was found.

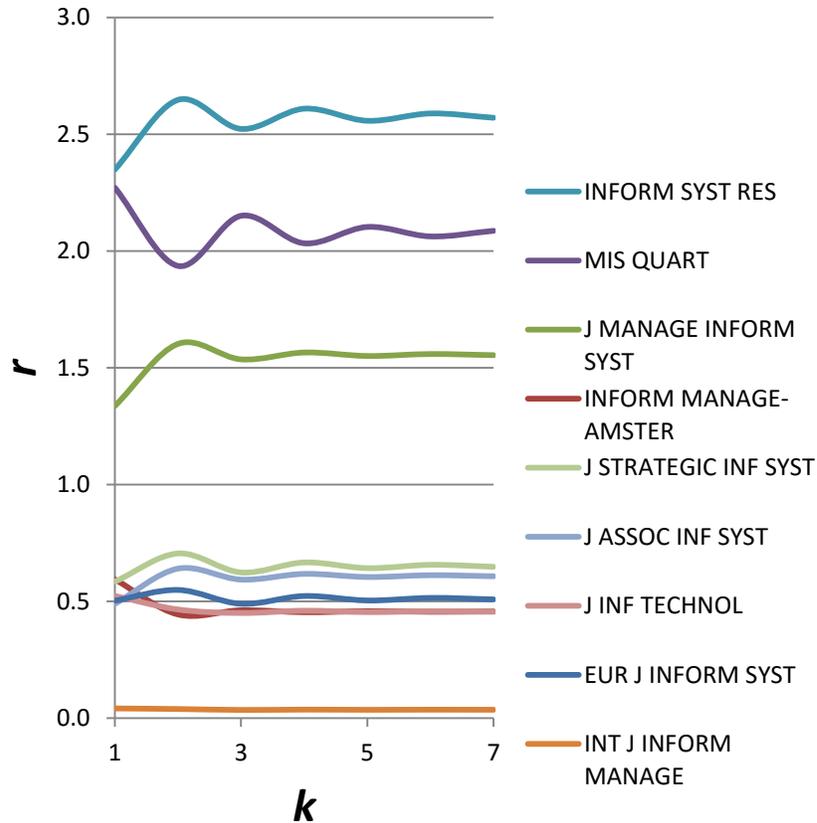

**Figure 3:** Convergence of PWR with iteration number $k$ for the nine MIS QUART+ journals for the case without self-citations.

But can the converged values of PWR also be considered as impact indicators of these journals? In our opinion, one can envisage three different options to interpret, for example, the results in Table 4:

(1) Since the authors of this paper are knowledgeable in information science (or scientometrics), the ranking of LIS journals can be interpreted on the basis of our professional experience. The rank-ordering of LIS journals by PWR could not be provided by us with an interpretation.

(2) Another way of interpreting the results would be to compare PWR with a most similarly designed journal metric. The SCImago Journal Rank (SJR), for example, uses an algorithm similar to PageRank; for the sake of comparison the values of SJR for these seven journals are included in Table 5.

**Table 5**: Seven strongly connected journals in LIS (JASIST+) ranked on their PWR within this group. For comparison, the SJR values from 2013 are included (see http://www.journalmetrics.com/values.php).

| Journal | PWR | SJR 2013 |
|---|---|---|
| *Inform Process Manag* | 1.79 | 0.751 |
| *JASIST* | 1.58 | 1.745 |
| *J Doc* | 1.48 | 0.876 |
| *J Inf Sci* | 1.25 | 1.008 |
| *Scientometrics* | 0.99 | 1.412 |
| *J Informetr* | 0.48 | 2.541 |
| *Inform Res* | 0.42 | 0.475 |

The columns for PWR and SJR correlate negatively with $r = –0.26$ (*n.s.*). This coefficient points out a weak relationship. Thus, both metrics measure different types of journal impact if they measure journal impact at all.

(3) A third way of interpreting the results is to compare the metric with an external criterion. For example, we could ask a sample of information scientists to assess the journals. However, we did not expect other assessments to differ from our own, and therefore did not pursue this option.

In sum, the indicator did not perform convincingly for journal ranking even in homogeneous sets.

**An inhomogeneous set**

Let us complete the analysis by combining the JASIST+ and MIS QUART+ sets into a single and arguably non-homogeneous set, since the one is from the LIS partition and the other from the MIS partition. Whereas the former set cites the latter generously, citations are not provided equally in the opposite direction.

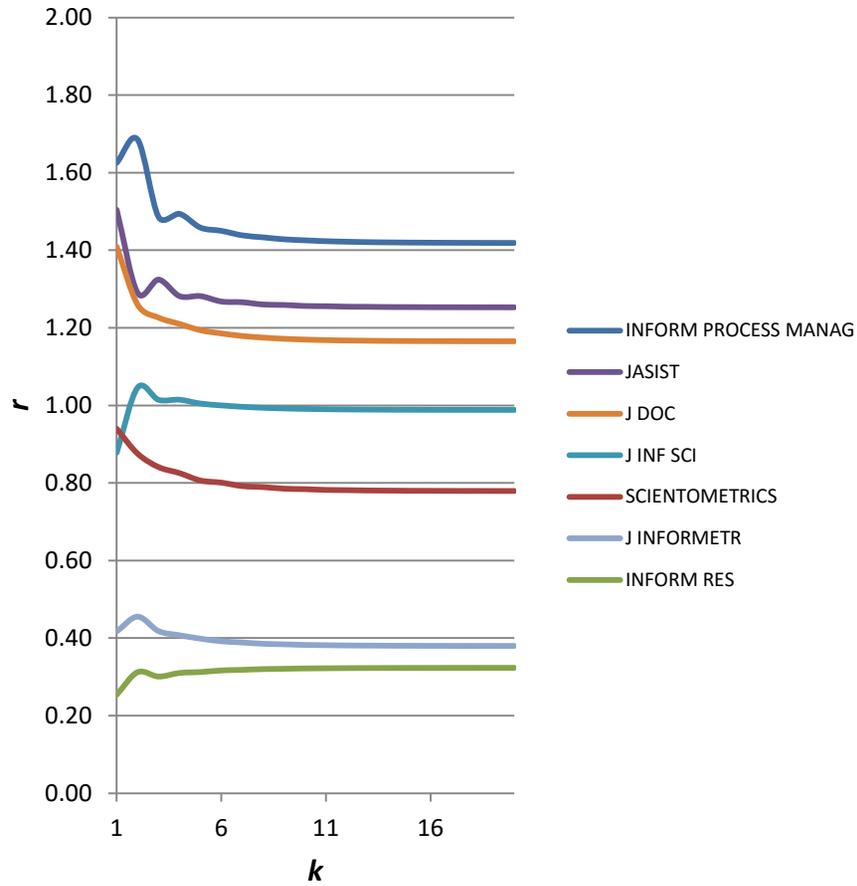 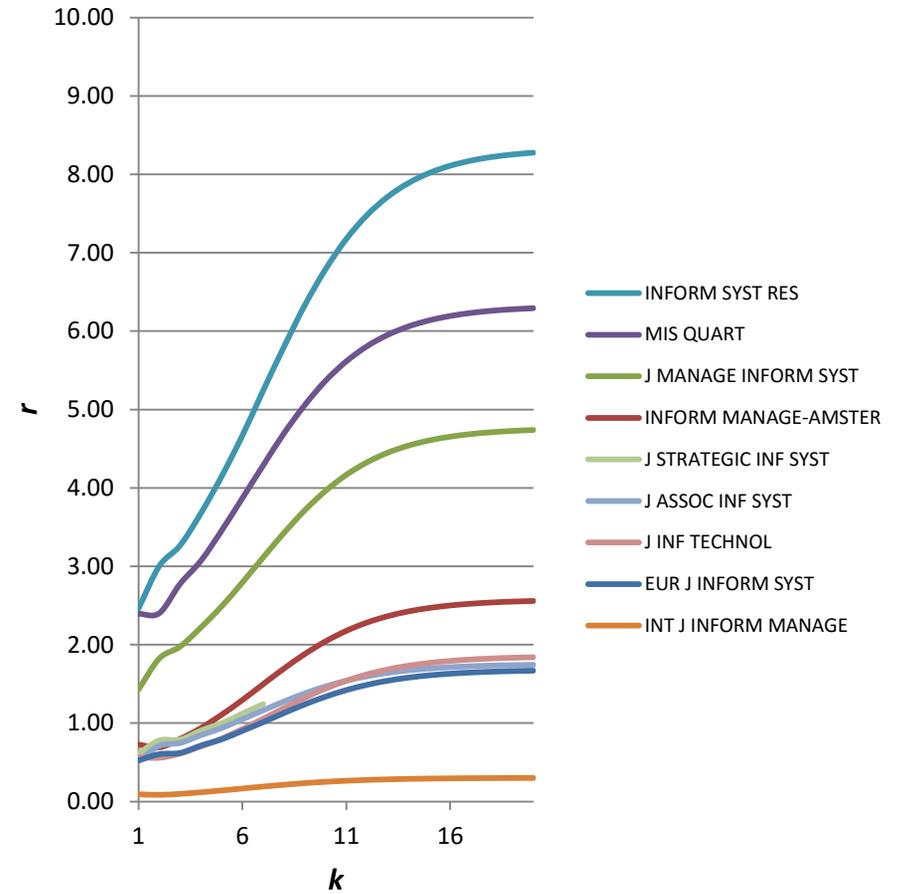

**Figure 4:** Convergence of PWR with iteration number $k$ for the seven JASIST+ journals within a heterogeneous environment (without self-citations).

**Figure 5:** Convergence of PWR with iteration number $k$ for the nine MIS QUART+ journals within a heterogeneous environment (without self-citations).

Figure 4 shows the convergence of PWR for the JASIST+ subgroup of journals. However, initial divergence of PWR at iteration number seven was noticed and final convergence was found for the MIS QUART+ journals only after 20 iterations in the case of a non-homogeneous set (Figure 5).[6] The difference between the two sets is illustrated by the two figures.

In other words, Ramanujacharyulu's PWR paradigm may offer a diagnostic tool for determining whether a journal set is homogeneous or not, but it may also fail to converge or to provide meaningful results in the case of heterogeneous sets. As noted, the application of PWR may have to be limited to strong components.

**Discussion and conclusions**

We investigated whether Ramanujacharyulu's (1964) metrics for power-weakness ratios could also be used as a meaningful indicator of journal status using the aggregated citation relations among journals. As noted, PWR was considered an attractive candidate for measuring journal impact because of its symmetrical handling of the rows and columns in the asymmetrical citation matrix, its recursive algorithm (which it shares with other journal indicators), and its mathematical elegance (Prathap & Nishy, in preparation). Ramanujacharyulu (1964) developed the algorithm for scoring tournaments (Prathap, 2014). However, journals compete in incomplete tournaments; in a round-robin tournament, all the teams are completely connected. If one team wins, the other loses; but this constraint is not valid for journals.

---

[6] After twenty iterations, the MIS QUART+ set also converged.

In order to be able to appreciate the results, we experimented with a subset of the Journal Citation Reports 2013: the 83 journals assigned to the WoS category "information and library science." One advantage of this subset is our familiarity with these journals, so that we were able to interpret empirical results (Leydesdorff and Bornmann, 2011 and 2016). Used as input into Pajek, the 83x83 citation matrix led to convergence, but not to interpretable results. Journals that are not represented on the "citing" dimension of the matrix—for example, because they no longer appear, but are still registered as "cited" (e.g., *ARIST*)—distort the PWR ranking because of their relatively low values in the denominator. However, when the not-citing journals were excluded from the top-15 ranking, the ranking still did not match our intuition about relative journal standing.

In a next step, we focused on two specific subsets, namely all the journals citing *JASIST* or *MIS Quart* one hundred times or more. These two relatively homogenous subsets converged easily and each provided a rank order. However, the Pearson correlation between PWR and SJR was negative ($r = -0.26$; *n.s.*) for the case of the seven LIS journals.

In summary, the indicator did not perform convincingly for journal ranking. This may also be due to the assumption of equal gain or loss when a citation is added on the cited or the citing side, respectively. Using PWR, journal *i* gains and journal *j* loses when a reference is added at location *ij*. However, as noted above, the association of "cited" with "power" and "citing" with "weakness" may be cultural. In our opinion, referencing is an actor category and can be studied in terms of behavior, whereas "citedness" is a property of a document with an expected dynamics very different from that of "citing" (Wouters, 1999).

In other words, the citation to Ramanujacharyulu (1964) is interesting and historically relevant to eigenvector centrality methods that predate Narin & Pinski (1976). However, the PWR method was conceived in 1964 as a way to evaluate round-robin tournaments, but "wins" and "losses" do not translate to citations. Citations have to be normalized because of field-specificity and the discussion of damping factors can also not be ignored since the transitivity among citations is not unlimited (Brin & Page, 1998). With this study, we have wished to show that a newly proposed indicator can be critically assessed.

**Acknowledgement**
The authors acknowledge Gangan Prathap for discussing the PWR method with us in detail. We are grateful to Thomson Reuters for providing the data of the Journal Citation Reports.

**References**
Bergstrom, C. (2007). Eigenfactor: Measuring the value and prestige of scholarly journals. *College & Research Libraries News*, 68, 314.
Blondel, V. D., Guillaume, J. L., Lambiotte, R., & Lefebvre, E. (2008). Fast unfolding of communities in large networks. *Journal of Statistical Mechanics: Theory and Experiment, 8*(10), P10008, 10001-10012.
Brin, S., & Page, L. (1998). The anatomy of a large-scale hypertextual Web search engine. *Computer Networks and ISDN Systems*, 30 (1-7), 107-117.
de Nooy, W., Mrvar, A., & Batagelj, V. (2011). Exploratory social network analysis with Pajek: Revised and expanded second edition. Cambridge: Cambridge University Press.Dong, S. B. (1977). A Block–Stodola eigensolution technique for large algebraic systems with non-symmetrical matrices. *International Journal for Numerical Methods in Engineering, 11*(2), 247-267.
De Visscher, A. (2010). An index to measure a scientist's specific impact. *Journal of the American Society for Information Science and Technology*, 61(2), 310–318.
De Visscher, A. (2011). What does the g-index really measure? *Journal of the American Society for Information Science and Technology*, 62(11), 2290–2293.
Dong, S. B. (1977). A Block–Stodola eigensolution technique for large algebraic systems with non-symmetrical matrices. *International Journal for Numerical Methods in Engineering, 11*(2), 247-267.
Franceschet, M. (2011). Page Rank: Standing on the shoulders of giants. *Communications of the ACM*, 54(6), 92-101.
Garfield, E., & Sher, I.H. (1963). New factors in the evaluation of scientific literature through citation indexing. *American Documentation*, 14, 195–201.

Gingras, Y., & Larivière, V. (2011). There are neither "king" nor "crown" in scientometrics: Comments on a supposed "alternative" method of normalization. *Journal of Informetrics, 5*(1), 226-227.

Guerrero-Bote, V. P., & Moya-Anegón, F. (2012). A further step forward in measuring journals' scientific prestige: The SJR2 indicator. *Journal of Informetrics, 6*(4), 674-688.

Kamada, T., & Kawai, S. (1989). An algorithm for drawing general undirected graphs. *Information Processing Letters, 31*(1), 7-15.

Kleinberg, J.M. (1999). Authoritative sources in a hyperlinked environment. *Journal of ACM*, *46* (5), 604–632.

Leydesdorff, L. (2006). Can Scientific Journals be Classified in Terms of Aggregated Journal-Journal Citation Relations using the Journal Citation Reports? *Journal of the American Society for Information Science & Technology, 57*(5), 601-613.

Leydesdorff, L. (2009). How are New Citation-Based Journal Indicators Adding to the Bibliometric Toolbox? *Journal of the American Society for Information Science and Technology*, 60(7), 1327–1336.

Leydesdorff, L., & Bornmann, L. (2012). Percentile Ranks and the Integrated Impact Indicator (*I3*). *Journal of the American Society for Information Science and Technology, 63*(9), 1901-1902.

Leydesdorff, L., & Bornmann, L. (2016). The Operationalization of "Fields" as WoS Subject Categories (WCs) in Evaluative Bibliometrics: The cases of "Library and Information Science" and "Science & Technology Studies". *Journal of the Association for Information Science and Technology, 67*(3), 707-714.

Leydesdorff, L., Bornmann, L., Mutz, R., & Opthof, T. (2011). Turning the tables in citation analysis one more time: Principles for comparing sets of documents *Journal of the American Society for Information Science and Technology, 62*(7), 1370-1381.

Milojević, S., & Leydesdorff, L. (2013). Information Metrics (iMetrics): A Research Specialty with a Socio-Cognitive Identity? *Scientometrics, 95*(1), 141-157.

Narin, F. (1976). *Evaluative Bibliometrics: The Use of Publication and Citation Analysis in the Evaluation of Scientific Activity*. Washington, DC: National Science Foundation.

Nicolaisen, J., & Frandsen, T. F. (2008). The Reference Return Ratio. *Journal of Informetrics, 2*(2), 128-135.

Opthof, T., & Leydesdorff, L. (2010). *Caveats* for the journal and field normalizations in the CWTS ("Leiden") evaluations of research performance. *Journal of Informetrics, 4*(3), 423-430.

Pinski, G., & Narin, F. (1976). Citation influence for journal aggregates of scientific publications: Theory, with application to the literature of physics. *Information Processing and Management*, 12(5), 297–312.

Prathap, G. (2014). The Best Team at IPL 2014 and EPL 2013-2014, *Science Reporter*, August 2014, 44-47.

Prathap, G. & Nishy, P. (in peparation). A size-independent journal impact metric based on social-network analysis. Preprint available at https://www.academia.edu/7765183/A_size-independent_journal_impact_metric_based_on_social-network_analysis

Price, D. J. de Solla (1976). A general theory of bibliometric and other cumulative advantage processes. *Journal of the American Society for Information Science, 27*(5), 292-306.

Price, D. J. d. Solla (1981). The Analysis of Square Matrices of Scientometric Transactions. *Scientometrics*, 3(1), 55-63.

Rafols, I., Leydesdorff, L., O'Hare, A., Nightingale, P., & Stirling, A. (2012). How journal rankings can suppress interdisciplinary research: A comparison between innovation studies and business & management. *Research Policy, 41*(7), 1262-1282.

Ramanujacharyulu, C. (1964). Analysis of preferential experiments. *Psychometrika*, 29(3), 257-261.

Todeschini, R., Grisoni, F., & Nembri, S. (2015). Weighted power–weakness ratio for multi-criteria decision making. *Chemometrics and Intelligent Laboratory Systems, 146*, 329-336.

Waltman, L., Yan, E., & van Eck, N. J. (2011a). A recursive field-normalized bibliometric performance indicator: An application to the field of library and information science. *Scientometrics, 89*(1), 301-314.

Waltman, L., Van Eck, N. J., Van Leeuwen, T. N., Visser, M. S., & Van Raan, A. F. J. (2011b). Towards a New Crown Indicator: Some Theoretical Considerations. *Journal of Informetrics, 5*(1), 37-47.

West, J. D., Bergstrom, T. C., & Bergstrom, C. T. (2010). The Eigenfactor metrics: A network approach to assessing scholarly journals. *College and Research Libraries*, 71(3), 236–244.

Wouters, P. (1999). *The Citation Culture*. Amsterdam: Unpublished Ph.D. Thesis, University of Amsterdam.

Yan, E., & Ding, Y. (2010). Weighted citation: An indicator of an article's prestige. *Journal of the American Society for Information Science and Technology*, 61(8), 1635–1643.

Yanovsky, V. (1981). Citation analysis significance of scientific journals. *Scientometrics, 3*(3), 223-233.

Zhirov, A., Zhirov, O., & Shepelyansky, D. L. (2010). Two-dimensional ranking of Wikipedia articles. *The European Physical Journal B, 77*(4), 523-531.
Price, D. J. d. Solla (1981). The Analysis of Square Matrices of Scientometric Transactions. *Scientometrics*, 3(1), 55-63.

Rafols, I., Leydesdorff, L., O'Hare, A., Nightingale, P., & Stirling, A. (2012). How journal rankings can suppress interdisciplinary research: A comparison between innovation studies and business & management. *Research Policy, 41*(7), 1262-1282.

Ramanujacharyulu, C. (1964). Analysis of preferential experiments. *Psychometrika*, 29(3), 257-261.

Todeschini, R., Grisoni, F., & Nembri, S. (2015). Weighted power–weakness ratio for multi-criteria decision making. *Chemometrics and Intelligent Laboratory Systems, 146*, 329-336.

Waltman, L., Yan, E., & van Eck, N. J. (2011a). A recursive field-normalized bibliometric performance indicator: An application to the field of library and information science. *Scientometrics, 89*(1), 301-314.

Waltman, L., Van Eck, N. J., Van Leeuwen, T. N., Visser, M. S., & Van Raan, A. F. J. (2011b). Towards a New Crown Indicator: Some Theoretical Considerations. *Journal of Informetrics, 5*(1), 37-47.

West, J. D., Bergstrom, T. C., & Bergstrom, C. T. (2010). The Eigenfactor metrics: A network approach to assessing scholarly journals. *College and Research Libraries*, 71(3), 236–244.

Wouters, P. (1999). *The Citation Culture*. Amsterdam: Unpublished Ph.D. Thesis, University of Amsterdam.

Yan, E., & Ding, Y. (2010). Weighted citation: An indicator of an article's prestige. *Journal of the American Society for Information Science and Technology*, 61(8), 1635–1643.

Yanovsky, V. (1981). Citation analysis significance of scientific journals. *Scientometrics, 3*(3), 223-233.

Zhirov, A., Zhirov, O., & Shepelyansky, D. L. (2010). Two-dimensional ranking of Wikipedia articles. *The European Physical Journal B, 77*(4), 523-531.

**Appendix 1**: PWR with Pajek

*Running a macro*

PWR can be calculated in Pajek using the Pajek macro PWR.mcr (available at http://www.leydesdorff.net/pwr/pwr.mcr). The macro calculates PWR for 19 iterations, hence for *k* up to and including 20.

The macro requires as its input a Pajek network file with arcs pointing from cited to citing journal. Select this network file in both the first and second network dropdown list. Ensure that the Options>Read – Write>Ignore Missing Values in menu Vector and Vectors is selected. If journals may have cited neither other journals nor itself, the Weakness score will be zero and the PWR will be calculated by dividing by zero. By default, Pajek uses 999999999 (the missing value) when a number is divided by zero. You may want to set this result to zero because the PWR is meaningless for journals that are not (self)citing. This option can also be set in the Options>Read-Write dialog, namely in the inut box after 'x / 0 =.' Finally run the macro with the Macro>Play command: select the file PWR.mcr.

The macro creates a lot of vectors: first the Power vectors, then the Weakness vectors, and finally the PWR vectors, labeled 'PWR at k = …' . The Pajek project file for the citation matrix 2013 for 83 journals in the category "information and library science" is provided as an example at http://www.leydesdorff.net/pwr/matrix83.paj .

*PWR calculation step by step*

If you want to restrict or expand the number of iterations, you may execute all steps one-by-one in Pajek instead of executing the macro.

Steps (in Pajek 3.15):
1. Multiply the original network by itself.
   * Select the original network both in the first and second network dropdown list first.
   * Execute the command Networks>Multiply Networks.
2. Repeat the multiplication with the original (power) network and the result of the previous multiplication *N* – 1 times (*N* the maximum number of iterations desired, counting the original network as iteration #0).
   * Select the Macro>Repeat Last Command command or press F10.
   * In the dialog screen select Fix (Second) Network. This is the original network.
   * Press Repeat Last Command.
   * Plug in the required total number of iterations minus 1.
   * Answer No (or Yes) to the question 'Write all intermediate reports to Report Window?'
3. Transpose the original network and repeat Steps 1 and 2 for the transposed (weakness) networks.
   * Select the original network in the top network dropdown list.
   * Apply the Network>Create New Network>Transform>Transpose 1-Mode command and answer Yes to the question Create a new Network as a result?
   * Select the transposed network also in the second network drop down list.
   * Execute the command Networks>Multiply Networks.
   * Select the Macro>Repeat Last Command command or press F10.
   * In the dialog screen select Fix (Second) Network. This is the original network.

* Press Repeat Last Command.
* Plug in the required total number of iterations minus 1.
* Answer No (or Yes) to the question 'Write all intermediate reports to Report Window?'
4. Calculate the row totals for all networks.
   * Select the original network in the top network dropdown list.
   * Use Network>Create Vector>Centrality>Weighted Degree>Output to obtain a vector of row sums for the original network.
   * Use the Macro>Repeat Last Command command or press F10 to repeat this for the other networks, use $2N + 1$ as the number of repetitions.
5. Calculate the quotient for each pair of matching power – weakness networks.
   * Ensure that missing values in vectors are disregarded by selecting this option in the Options>Read - Write menu.
   * Select the vector of row counts of the original (power) network in the first vector dropdown list and the vector for the first weakness network (= transposed original network) in the second vector dropdown list.
   * Execute the Vectors>Divide (First/Second) command to create a new vector with the PWR scores for the original network.
   * Again, select the original (power) network in the first vector dropdown list (the first weakness network should still be selected in the second vector dropdown list).
   * Use the Macro>Repeat Last Command command or press F10 to repeat this for the other networks, use $N$ as the number of repetitions.
   NOTE: if interest is restricted in the PWR scores of the last iteration, it suffices to divide the power and weakness vectors of the last iteration.
6. Inspect the resulting vectors (File>Vector>View/Edit or Vector>Info) or send them to other software with commands in the Tools menu.

**Appendix 2: PWR with Excel**

1. As examples two files are provided at http://www.leydesdorff.net/pwr/jasist.xlsx and http://www.leydesdorff.net/pwr/mmult.xlsx, respectively. Both files contain the 7x7 cited-citing matrix in array (B4:H10).
   a. JASIST.xlsx is based on matrix multiplication using the formulas of linear algebra;
   b. MMULT.xlsx uses the mmult() function in Excel.
2. The first matrix multiplication multiplies each row of this matrix with the start vector (I4:I10), taken as a vector with each element having the value 1/n (n = 7 in this case). This is actually the raw count of citations, except for the factor of 7, and is kept at (J4:J10). The new eigenvector is obtained at column K by normalizing this. The multiplication is then done repeatedly. At the end of the $k^{th}$ cycle one obtains the power vector p(k).
3. The weakness iteration repeats this with the transposed matrix. One obtains the weakness vector w(k) at the end of the $k^{th}$ cycle.
4. The power-weakness ratio r is then given by r(k) = p(k)/w(k) at the end of the k cycles.
5. In sheet 2, the whole calculation is repeated for the case without self-citations.
6. The MMULT function returns **#VALUE!** if the output exceeds 5460 cells (n ≤ 73); see at https://support.microsoft.com/kb/166342?wa=wsignin1.0